\documentclass[aps,tightenlines,14pt,showpacs]{revtex4}

\usepackage{graphicx}  
\usepackage{bm}

\begin{document}

\title{Effects of nuclear fragmentation on single particle and collective
motions at low energy\footnote{Proceedings of the ``10$^{\rm th}$
international conference on nuclear reaction mechanisms'',
Varenna, 9-13 June 2003.}}
\author{C.~Barbieri}
  \email{barbieri@triumf.ca}
  \homepage{http://www.triumf.ca/people/barbieri}
  \affiliation
     {TRIUMF, 4004 Wesbrook Mall, Vancouver, 
          British Columbia, Canada V6T 2A3 \\  }
\author{W.~H.~Dickhoff}
  \email{wimd@wuphys.wustl.edu}
  \homepage{http://www.physics.wustl.edu/~wimd}
  \affiliation
     {Department of Physics, Washington University,
	 St.Louis, Missouri 63130, USA \\  }

\preprint{TRI-PP-03-26}

\begin{abstract}
 The Faddeev technique is employed to study the influence of both
particle-particle and particle-hole phonons on the  one-hole spectral
function of ${}^{16}{\rm O}$.
 The formalism includes the effects of nuclear fragmentation and 
 accounts for collective excitations at a random phase approximation (RPA)
level. The latter are subsequently summed to all orders by the Faddeev
equations to obtain the nucleon self-energy.
The results indicate that the characteristics of hole fragmentation
are related to the low-lying states of ${}^{16}{\rm O}$ and an improvement
of the description of this spectrum is required to understand the details
of the experimental strength distribution.
 A first calculation to improve on this includes
the mixing of two-phonon configuration with particle-hole ones. It is seen
that this allows to understand the formation of excited states which cannot
be described by the RPA approach.
\end{abstract}

\maketitle

\large

\section{Introduction}

The studies of spectral distribution in nuclei during the last decade 
has made available the best information to date regarding correlations in
nuclear systems~\cite{Sick97}.
Experimentally, $(e,e'p)$ reactions have provided precise information for
spectroscopic factors of states close to the Fermi energy~\cite{Louk,leus}
in finite nuclei. Moreover, information on the
tail of the spectral distribution at high missing energies and momenta, 
generated by short-range correlations (SRC), has recently become
available~\cite{DanielaLund}.
The observed spectroscopic factors have been reproduced
for selected nuclei,
such as ${}^{48}{\rm Ca}$ and ${}^{90}{\rm Zr}$~\cite{CaZr}, and a
particularly successful comparison between theory and experiment
has been obtained for ${}^{7}{\rm Li}$~\cite{LapLi7}. In all these
calculations, the effects of configuration mixing at low energy was taken
properly into account.   Besides, the effects of SRC have been investigated
for different systems and are known to give a 
constant reduction of the spectroscopic strength of 10-15\%, for the
fragments close to the Fermi energy~\cite{pmd94-95,jastO16,bv91}.

For the case of ${}^{16}{\rm O}$, there still exists a substantial 
disagreement between experiment~\cite{leus} and
theory~\cite{pmd94-95,jastO16,muether2h1pO16,GeurtsO16,FaddRPAO16}.
The latter has been able to explain only a part of the quenching of
spectroscopic factors for proton knockout from the $p$-shell.  The results
of Refs.~\cite{GeurtsO16,muether2h1pO16} show that the reasons
for this discrepancy
have to be looked for in terms of long-range correlations (LRC) and in
particular in the couplings of single-particle (sp) motion to
low-energy collective excitations.

In this paper we report about the work done along this line in
Refs.~\cite{FaddRPAO16,2phERPAO16} in order to approach the above issues
for ${}^{16}{\rm O}$. Sec.~\ref{sec:2} describes the formalism employed to 
describe the coupling of particle-hole (ph), particle-particle (pp) and
hole-hole (hh) collective excitation to the sp propagator. This is done
by embedding the technique of Faddeev equations into the self-consistent
Green's function formalism~\cite{FaddRPA1,BaDLund}.  The results of this
calculation are given in Sec.~\ref{sec:3}, where the effects of nuclear
fragmentations and the random phase approximation (RPA) in the description
of phonons are discussed. These calculations show that a proper description
of the experimental spectral strength requires an improvement
of the spectrum, that goes beyond the RPA approach.
 A first step in this direction is reported in Sec.~\ref{sec:4}, where the
Bethe-Salpeter equation (BSE) is employed to pursue the improvement of
the ph-RPA approach.

\section{Faddeev approach the calculation of the single-particle
       Green's function}
\label{sec:2}

\begin{figure}[t]
 \begin{center}
 \includegraphics[height=1.8in]{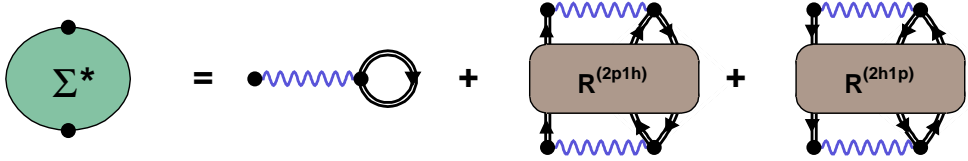}
 \end{center}
\vspace{-0.5in}
\caption[]{\small
 Diagrams contributing to the irreducible self-energy $\Sigma^*$.
   The double lines represent a dressed propagator and 
   the wavy lines correspond to G-matrix interactions.  The first term is the 
   Brueckner-Hartree-Fock potential while the others represent the 2p1h/2h1p 
   or higher contributions that are approximated through the Faddeev TDA/RPA
   equations.
\label{fig:selfenergy} }
\end{figure}

We consider the calculation of the sp Green's function
\begin{equation}
 g_{\alpha \beta}(\omega) ~=~ 
 \sum_n  \frac{ \left( {\cal X}^{n}_{\alpha} \right)^* \;{\cal X}^{n}_{\beta} }
                       {\omega - \varepsilon^{+}_n + i \eta }  ~+~
 \sum_k \frac{ {\cal Y}^{k}_{\alpha} \; \left( {\cal Y}^{k}_{\beta} \right)^* }
                       {\omega - \varepsilon^{-}_k - i \eta } \; ,
\label{eq:g1}
\end{equation}
from which both the one-hole and one-particle spectral functions, for the
removal and addition of a nucleon, can be extracted.
In Eq.~(\ref{eq:g1}),
${\cal X}^{n}_{\alpha} = {\mbox{$\langle {\Psi^{A+1}_n} \vert $}}
 c^{\dag}_\alpha {\mbox{$\vert {\Psi^A_0} \rangle$}}$%
~(${\cal Y}^{k}_{\alpha} = {\mbox{$\langle {\Psi^{A-1}_k} \vert $}}
 c_\alpha {\mbox{$\vert {\Psi^A_0} \rangle$}}$) are the
spectroscopic amplitudes for the excited states of a system with
$A+1$~($A-1$) particles and the poles $\varepsilon^{+}_n = E^{A+1}_n - E^A_0$%
~($\varepsilon^{-}_k = E^A_0 - E^{A-1}_k$) correspond to the excitation
energies with respect to the $A$-body ground state.
The one-body Green's function can be
computed by solving the Dyson equation
\begin{equation}
 g_{\alpha \beta}(\omega) =  g^{0}_{\alpha \beta}(\omega) \; +  \;
   \sum_{\gamma \delta}  g^{0}_{\alpha \gamma}(\omega) 
     \Sigma^*_{\gamma \delta}(\omega)   g_{\delta \beta}(\omega) \; \; ,
\label{eq:Dys}
\end{equation}
where the irreducible self-energy $\Sigma^*_{\gamma \delta}(\omega)$ acts
as an effective, energy-dependent, potential. The latter can be expanded in a
Feynman-Dyson series~\cite{fetwa,AAA} in terms the exact
propagator $g_{\alpha \beta}(\omega)$, which itself is a solution
of Eq.~(\ref{eq:Dys}).
 In this expansion, $\Sigma^*_{\gamma  \delta}(\omega)$ can be represented
as the sum
of a one-body Hartree-Fock potential and terms that describe the coupling
between the sp motion and more complex excitations~\cite{Win72}.
 This separation is depicted by the diagrams of Fig.~\ref{fig:selfenergy},
where the last two diagrams are expressed in terms of
the three-line irreducible propagator $R(\omega)$ which describes
the propagation of at least 2h1p or 2p1h at the same time.
 It is at the level of $R(\omega)$ that the correlations involving
interactions between different collective modes have to be included.

 The SCGF approach consist of first solving the self-energy and the Dyson
Eq.~(\ref{eq:Dys}) in terms of a mean-field IPM
propagator $g^{0}_{\alpha , \beta}(\omega)$. The (dressed) solution 
$g_{\alpha \beta}(\omega)$ is then used to evaluate the an improved 
self-energy, which then contains the effects of fragmentation. The whole
procedure is iterated until self-consistency is reached.
Baym and Kadanoff showed that a self-consistent solution of the
above equation guarantees the fulfillment of the principal
conservation laws~\cite{bk61}.


\subsection{Faddeev approach to the self-energy}
\label{sec:2fadd}

 In the following we are interested in describing the coupling of
sp motion to ph and pp(hh) collective excitations of the system.
All the relevant information regarding the latter are included in
the Lehmann representations of the polarization propagator
\begin{eqnarray}
 \Pi_{\alpha \beta , \gamma \delta}(\omega) &=& 
 \sum_{n \ne 0}  \frac{  {\mbox{$\langle {\Psi^A_0} \vert $}}
            c^{\dag}_\beta c_\alpha {\mbox{$\vert {\Psi^A_n} \rangle$}} \;
             {\mbox{$\langle {\Psi^A_n} \vert $}}
            c^{\dag}_\gamma c_\delta {\mbox{$\vert {\Psi^A_0} \rangle$}} }
            {\omega - \left( E^A_n - E^A_0 \right) + i \eta } 
\nonumber \\
 &-& \sum_{n \ne 0} \frac{  {\mbox{$\langle {\Psi^A_0} \vert $}}
              c^{\dag}_\gamma c_\delta {\mbox{$\vert {\Psi^A_n} \rangle$}} \;
                 {\mbox{$\langle {\Psi^A_n} \vert $}}
             c^{\dag}_\beta c_\alpha {\mbox{$\vert {\Psi^A_0} \rangle$}} }
            {\omega - \left( E^A_0 - E^A_n \right) - i \eta } \; ,
\label{eq:Pi}
\end{eqnarray}
that describes the excited states in the system  with $A$ particles,
and of the two-particle propagator that is relevant for the $A \pm 2$
excitations
\begin{eqnarray}
 g^{II}_{\alpha \beta , \gamma \delta}(\omega) &=& 
 \sum_n  \frac{  {\mbox{$\langle {\Psi^A_0} \vert $}}
                c_\beta c_\alpha {\mbox{$\vert {\Psi^{A+2}_n} \rangle$}} \;
                 {\mbox{$\langle {\Psi^{A+2}_n} \vert $}}
         c^{\dag}_\gamma c^{\dag}_\delta {\mbox{$\vert {\Psi^A_0} \rangle$}} }
            {\omega - \left( E^{A+2}_n - E^A_0 \right) + i \eta }
\nonumber \\  
&-& \sum_k  \frac{  {\mbox{$\langle {\Psi^A_0} \vert $}}
    c^{\dag}_\gamma c^{\dag}_\delta {\mbox{$\vert {\Psi^{A-2}_k} \rangle$}} \;
                 {\mbox{$\langle {\Psi^{A-2}_k} \vert $}}
                  c_\beta c_\alpha {\mbox{$\vert {\Psi^A_0} \rangle$}} }
            {\omega - \left( E^A_0 - E^{A-2}_k \right) - i \eta } \; .
\label{eq:g2}
\end{eqnarray}
\begin{figure}
 \begin{center}
    \parbox[b]{.5\linewidth}{
\caption[]{Example of diagrams that are summed to all orders by means of 
the Faddeev equations.
 \vspace{1in} 
\label{fig:FaddSum} }  } 
    \parbox[b]{7mm}{~}
    \parbox[b]{.3\linewidth}{
 \includegraphics[height=2in]{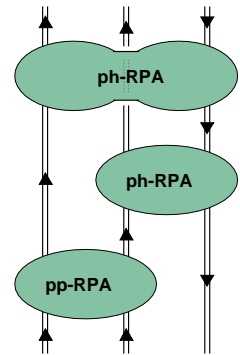} }
\end{center}
\end{figure}
 In the calculation of Sec.~\ref{sec:3}, these quantities
have been evaluated by solving the dressed TDA/RPA (DTDA/DRPA) equations
~\cite{schuckbook,DRPAg_papers} for the ph and pp case, respectively.
The fact that a dressed sp propagator is used brings in to the calculation
the effects of the strength distribution for particle and hole fragments.
One must notice that the propagators~(\ref{eq:Pi}) and~(\ref{eq:g2})
are in fact the solution of more general Bethe-Salpeter equations,
that also include correlations beyond the RPA level.
One of these is the mixing of many-phonons in the kernel of the BSE.
A first calculation in this direction
has been done for the ph propagator $\Pi(\omega)$~\cite{2phERPAO16}
and is reported in Sec.~\ref{sec:4}.

Once one has obtained the ph~(\ref{eq:Pi}) and pp(hh)~(\ref{eq:g2})
propagators, their coupling to the sp motion has to be accounted for
in the nuclear self-energy. To do this the 2p1h and 2h1p propagators
of Fig.~\ref{fig:selfenergy} are computed by solving a set of Faddeev
equations~\cite{Fadd1,gloeck}.
 The details of the formalism employed in this calculation have been
discussed in
Ref.~\cite{FaddRPA1} and will not be shown here. For the present
discussion
it is sufficient to note that this approach computes the motion of
three-quasiparticle excitations in the same way it is normally done
to solve the three-body problem.
 This allows to include  the effects of ph and pp(hh) motion not only
at the TDA level but also with more collective RPA phonons and beyond.
 Such excitations are coupled to each other by the
Faddeev equations, generating  an infinite series of diagrams like the
one shown in Fig.~\ref{fig:FaddSum}. Moreover, Pauli correlations are
properly taken into account at the 2p1h/2h1p level.

\section{Results for the single-particle spectral function 
          of ${}^{16}{\rm O}$}
\label{sec:3}

 In the calculations described below, the Dyson equation was solved within
a model space consisting of harmonic oscillator sp states.
An oscillator parameter $b =$ 1.76~fm was chosen (corresponding to
$\hbar\omega =$ 13.4~MeV) and all the first four major
shells (from $1s$ to $2p1f$) plus the $1g_{9/2}$ where included.
 Inside this model space, the interaction used was a Brueckner
G-matrix~\cite{CALGM} derived from the Bonn-C potential~\cite{bonnc}.
The results of Refs.~\cite{CaZr,2phERPAO16}, suggest that
this model space is large enough to properly account
for the low-energy collective states in which we
are mainly interested in, the SRC being accounted for through the G-matrix
effective interaction.
 We note that energy dependence of the G-matrix 
also contributes to the normalization of the spectral strength.
 In Ref.~\cite{GeurtsO16}, it was shown that this accounts
for the extra depletion of spectroscopic factors that is induced by SRC,
at least for the normally occupied shells in the IPM.

We mention that the solution of RPA equations may lead to collective
instabilities whenever the iteration becomes too attractive.
A particular sensitivity to the strength of the G-matrix interaction
was found in the calculation of~\cite{FaddRPAO16}, only for the lowest
isoscalar $0^+$ solution. For this reason it was chosen to artificially
shift its eigenvalue to the experimental energy of
first excited state, 6.05~MeV, during the successive iterations.
 Note that the screening due the redistribution of the sp strength tends
to reduce the instability problem in DRPA.
Moreover, the origin of the instability for this particular solution
has been identified and corrected later, in Ref.~\cite{2phERPAO16}.

\begin{table}[t]
 \begin{center}
 \begin{tabular}{ccccccccccccc}
  \hline 
  \hline 
    Shell      & ~~ & TDA    &~~ & RPA    & ~~& 1st itr. &~ ~& 2nd itr. &~~ & 3rd
    itr. &~ & 4th itr. \\
  \hline 
 $Z_{p_{1/2}}$ & & 0.775  & & 0.745  & & 0.775    & &  0.777   & & 0.774    & & 0.776 \\ \\
 $Z_{p_{3/2}}$ & & 0.766  & & 0.725  & & 0.725    & &  0.727   & & 0.722    & & 0.724    \\
               & &        & &        & & 0.015    & &  0.027   & & 0.026    & & 0.026 \\ \\
  \hline  \\
%
%
%
  Total occ.   & & 14.56  & & 14.56  & & 14.56    & &  14.57   & & 14.58    & & 14.63   \\
  \hline 
  \hline 
 \end{tabular}
  \end{center}
    \parbox[t]{1.\linewidth}{
      \caption[]{\label{tab:SpectFact}
  Hole spectroscopic factors ($Z_{\alpha}$) for knockout of a $\ell =1$
      proton from ${}^{16}{\rm O}$.
       These results refer to the initial IPM calculation and the first
      four iterations of the DRPA equations.
       All the values are given as a fraction of the corresponding IPM value.
       Also included is the total number of nucleons, as deduced from the
       complete one-hole spectral function, for each iteration.}
      }
\end{table}

\subsection{Effects of RPA correlations and fragmentation}

 For an IPM ansatz the TDA calculation is equivalent to the one of 
Ref.~\cite{GeurtsO16} and yields the same result for the spectroscopic 
factors of the main particle and hole shells
close to the Fermi energy.
  These correspond to 0.775 for $p_{1/2}$ and 0.766
for $p_{3/2}$ and are reported in Table~\ref{tab:SpectFact}.
 The introduction of RPA correlations reduces these values and brings them
down to 0.745 and 0.725, respectively.
This result reduces the discrepancy with the experiment by about 4\% and
shows that collectivity beyond the TDA level is relevant to
explain the quenching of spectroscopic factors.
 Since the present formalism does not account for center-of-mass
effects, the above quantities need to be increased by about 7\%
before they are compared with the experiment~\cite{O16com}.

\begin{figure}
 \begin{center}
 \includegraphics[height=4in]{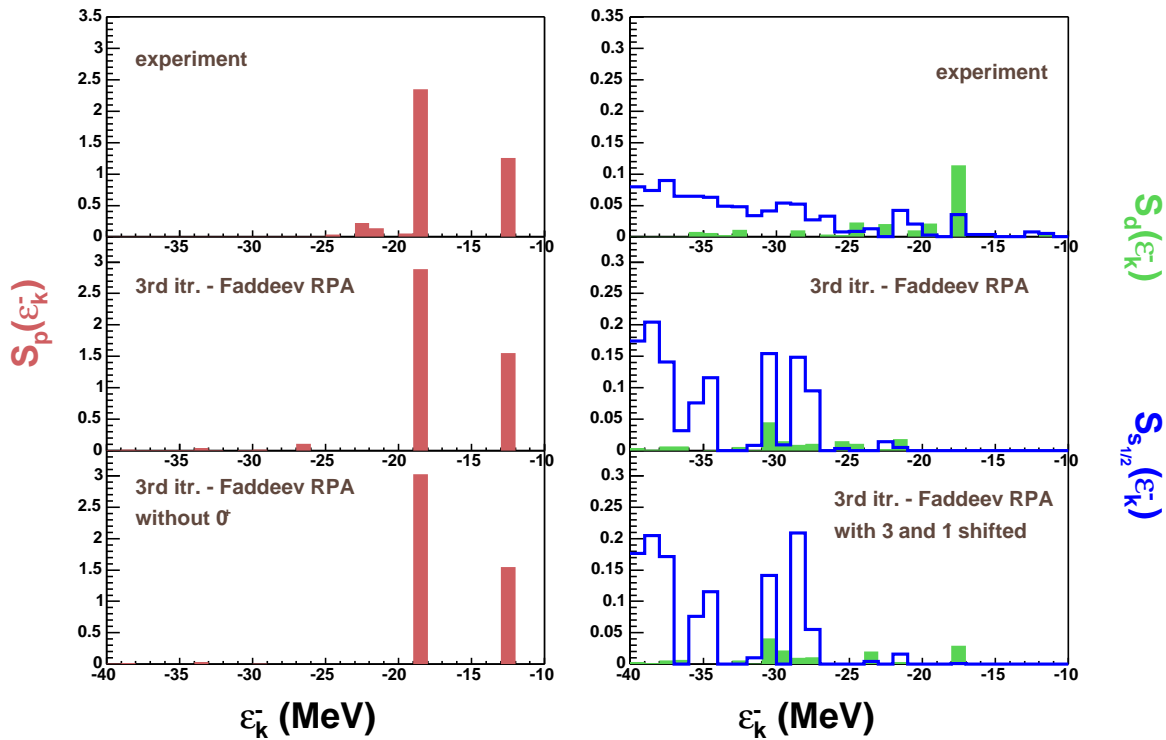}
 \end{center}
\caption[]{One-proton removal strength as a function of the 
   hole sp energy $\varepsilon^{-}_k = E^A_0 - E^{A-1}_k$
    for ${}^{16}{\rm O}$ for angular momentum
  $\ell =1$ (left) and $\ell =0,2$ (right).
   For the positive parity
   states, the solid bars correspond
   to results for $d_{5/2}$ and $d_{3/2}$ orbitals, while the
   thick lines refer to $s_{1/2}$.  
   The top panels show the experimental values taken from~\cite{leus}.
   The mid panels give the theoretical results for the self-consitent
   spectral function.
   The bottom panels show the results obtained by repeating the 3$^{\rm rd}$
   iteration with a modified ph-DRPA spectrum, in which
   the lowest eigenstates have been shifted to the corresponding
   experimental values.
\label{fig:p+sd_hsf} }
\end{figure}

The RPA results were iterated a few times, with the aim of studying
the effects of fragmentation on the RPA phonons and, subsequently,
on the spectral strength. 
Since we are looking at low-energy excitations,
it is sufficient to keep track only of the most occupied
fragments that appear ---close to the Fermi energy--- in the
(dressed) sp propagator, Eq.~(\ref{eq:g1}),
while the residual strength is collected in a effective
pole~\cite{jyuan,FaddRPAO16}.
As shown in Table~\ref{tab:SpectFact}, only 
a few iterations are required to reach convergence. The converged 
distribution of one-hole strength is compared in with the experimental one
in Fig.~\ref{fig:p+sd_hsf}. 
 The main difference between these results and the one obtained by using
an IPM input is the appearance of a second smaller $p_{3/2}$ fragment at 
-26.3~MeV. This peak arises in the first two iterations and appears
to become stable in the last one, with a spectroscopic factor of 2.6\%.
 This can be interpreted as a peak that describes the fragments seen
experimentally at slightly higher missing energy.
  This is the first time that such a fragment is
obtained in calculations of the spectral strength.
Further insight into the appearance of this strength is discussed
in Sec.~\ref{sec:hsf_vs_O16}.

 A second effect of including fragmentation in the construction of the RPA 
phonons is to increase the strength of the main hole
peaks.
  The $p_{1/2}$ 
strength increases from the 0.745 obtained with IPM input to 0.776,
essentially canceling the improvement gained by 
the introduction of RPA correlations over the TDA ones.
 The main peak of the $p_{3/2}$ remains at 0.722 but the appearance of the
secondary fragment slightly increases the overall strength at low
energy as well.
 This behavior is due to the fact that redistribution of the 
strength (and principally, the reduction of the main spectroscopic
factors) tends to screen the nuclear interaction, with respect to
the IPM case.
%
%
 Obviously, this makes the disagreement with experiments a little worse
and additional work is needed to resolve the disagreement with the data.
Nevertheless it is clear 
that fragmentation is a relevant feature of nuclear systems and that
it has to be properly taken into account.

 Together with the main fragments, the Dyson equation produces also a
large number of solutions with small spectroscopic factors.
 This strength extends down to about -130~MeV for the one-hole spectral
function and up to about 100~MeV for the one-particle case. 
This background represents 
the strength that is removed from the main peaks
and shifted up to medium missing energies.
 Another 10\% of strength is moved to very high energies
due to SRC~\cite{bv91}. Its location cannot be explicitly
calculated in the present approach but the effects on the
reduction of spectroscopic factors at low energy is accounted
through the
energy dependence of the G-matrix.
Accordingly the total number of 
nucleons deduced from the fully self-consistent spectral function is
about a 10\% lower than $A =$~16, as reported in Table~\ref{tab:SpectFact}.

 We note that the present approach gives predictions also for the strength
distribution of quasiparticle excitations. Although little experimental
information is available for the addition of a nucleon (as well as for
the removal of a neutron).
 Recently, knockout and nucleon transfer reactions, employing nuclear probes,
have been applied to measure the hole spectroscopic factors of both stable
and exotic nuclei~\cite{loukd3He,Hansen}.  These tools can now reach better
accuracies than in the past and may, in principle, be used to extract
information on the one-particle spectral distribution.

\subsection{Role of $0^+$ and $3^-$ excited states in ${}^{16}{\rm O}$}
\label{sec:hsf_vs_O16}

 A deeper insight into the mechanisms that generate the fragmentation
pattern can be
gained by investigating directly the connection between
the spectral function and some specific collective states.
 To clarify this point we repeated the third iteration using exactly the same
input but without keeping the troublesome ph $0^+$ state at its 
experimental energy, it was discarded instead.
 The resulting $p$ hole spectral function is shown in the lower-left panel
of Fig.~\ref{fig:p+sd_hsf}.
 In this calculation no breaking of the $p_{3/2}$ strength is obtained but 
only a single peak is found with a spectroscopic factor equal to 0.75, which
is the sum of the two fragments that are obtained when the 
$0_2^+$ state is taken into account.
 This result can be interpreted by considering the $p_{3/2}$ fragments as a
hole in the ground state and in the first excited $0^+$ state of the
${}^{16}{\rm O}$ core, respectively.
 The correct fragmentation pattern can be reproduced only when latter two levels
are close enough, in energy, to one another.

The other two low-lying states of ${}^{16}{\rm O}$ that may be of some
relevance are the isoscalar $1^-$ and $3^-$. 
These excitations are reproduced reasonably well
by RPA type calculations~\cite{czer} but are typically found at
$\sim$3~MeV above the experimental results.
 The lower panels of Fig.~\ref{fig:p+sd_hsf} show the results for the 
even parity spectral functions that are obtained when both
the $3^-$ and the $1^-$ ph-DRPA solutions are shifted to
match their experimental values. 
In this case, a $d_{5/2}$ hole peak is obtained at a missing energy
-17.7~MeV, in agreement with 
experiment,  and with a spectroscopic strength of~0.5\%.
%

It should be noted that many {\em weak} fragments (i.e. relative to the
addition or knockout for a nucleon with a  small
spectroscopic factor) are seen throughout the nuclear table of isotopes.
 Most of these
have no direct explanation in terms of mean-field sp states and 
contain more complicated components, involving 2p1h/2h1p configurations
and beyond.
Jennings and Escher have shown that different definitions of the
overlap wave function can be given that completely describe
the sp motion inside the nuclear medium. 
 Although, each of them sample the Pauli
correlations in a different way, resulting in very different
normalizations of the amplitudes of weak states~\cite{Byron02}. 
This may have important
consequences for the reliability of spectroscopic factors extracted from
proton emission and pick-up/stripping reactions~\cite{Jim}.

\section{Extension of RPA and spectrum of ${}^{16}{\rm O}$}
\label{sec:4}

\begin{figure}
 \begin{center}
    \parbox[b]{.5\linewidth}{
  \caption[]{\small
 Examples of contributions involving the coupling of two independent
ph phonons. 
 In total, there are sixteen possible diagrams of this type, obtained
by considering all the possible couplings to a ph state.
The two-phonon ERPA equations sum all of these
contributions in terms of dressed sp propagators.
\label{fig:2PiERPA} } }
    \parbox[b]{7mm}{~}
    \parbox[b]{.45\linewidth}{
 \includegraphics[width=.8\linewidth]{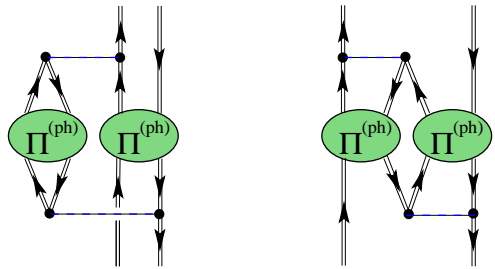} }
 \end{center}
\end{figure}

The above results suggest that the main impediment for further improvements
of the description of the experimental spectral strength is associated
with the deficiencies of the RPA (DRPA) description of the excited states.
One important problem
is the appearance of at most one collective phonon for a given $J^{\pi},T$
combination while several low-lying isoscalar $0^+$ and $2^+$
excited states are observed at low energy in ${}^{16}{\rm O}$, as well
as additional $3^-$ and $1^-$ states.
A possible way to proceed would be to first concentrate on an improved 
description of the collective phonons by extending the RPA to explicitly
include the coupling to two-particle$-$two-hole (2p2h) states.
Such an extended RPA procedure has been applied in heavier nuclei with
considerable success~\cite{CaZr}.
In order to be relevant for ${}^{16}{\rm O}$, this approach requires
an extension in which the coherence of the 2p2h states is included
in the form of the presence of two-phonon excitations~\cite{wim2ph}.
This correspond to  adding  two-phonon configurations to the kernel
of the BSE for the ph propagator~(\ref{eq:Pi}),
as  shown in Fig.~\ref{fig:2PiERPA}~\cite{2phERPAO16}. 

 We note that there has been a tremendous progress in recent years in
the microscopic description of {\em p} shell nuclei using Green's Function
Monte Carlo and no-core shell model methods~\cite{GFMC,NoCshell}.
 A possible application of the no-core shell model to ${}^{16}\textrm{O}$
would properly include the 4p4h effects relevant for describing
the spectrum of ${}^{16}{\rm O}$~\cite{O16_0+}. However the description of
spectroscopic
factors will require the construction of effective operators to 
include the effects of short-range correlations on these quantities
whereas they are automatically included in the SCGF method. Whence the
interest in studying the spectrum of ${}^{16}{\rm O}$ with
the present method.

\begin{figure}[t]
 \begin{center}
    \parbox[b]{.55\linewidth}{
 \includegraphics[width=1.\linewidth]{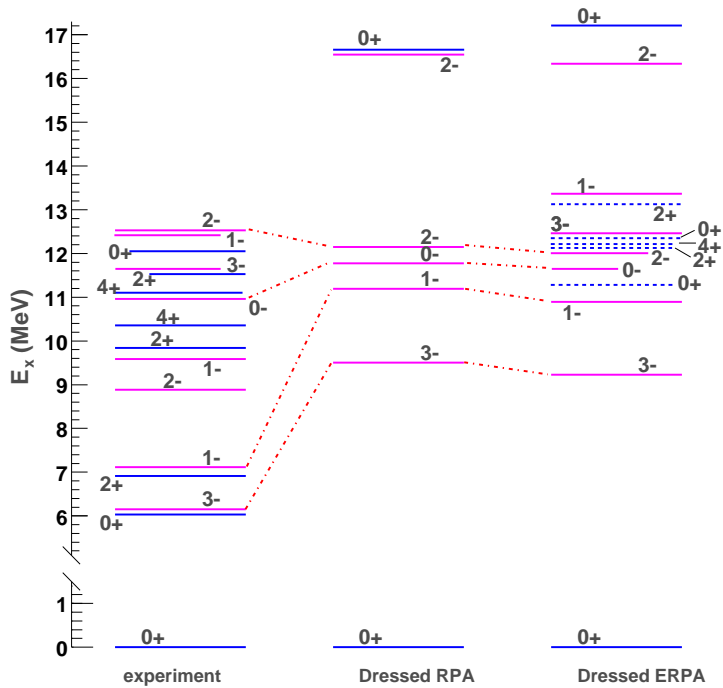} }
    \parbox[b]{7mm}{~}
    \parbox[b]{.40\linewidth}{
\caption[Two-phonon ERPA results for ${}^{16}\textrm{O}$; dressed input]
 {\small 
 Results for the DRPA and the two-phonon ERPA propagator of ${}^{16}\textrm{O}$
 with the dressed input propagator computed in Sec.~\ref{sec:3}, middle and
 last column respectively.
  In solving the ERPA equation, the lowest $3^-$, $1^-$ and $0^+$ levels
 of the DRPA propagator where shifted to their experimental energies. All
 other DRPA solutions were left unchanged.
 The excited states indicated by dashed lines are those for which the (E)RPA
 equation predicts a total spectral strength $Z_{n_\pi}$ lower than 10\%.
  The first column reports the experimental results~\cite{azj}.
\label{fig:2PiERPA_3g} }
 \vspace{1.in} }
 \end{center}
\end{figure}

The formalism to include two-phonon configurations has been presented
in Ref.~\cite{2phERPAO16}, where it is referred as two-phonon extended RPA
(ERPA).
In this work the ph-DRPA equation has been solved first,
using the self-consistent sp propagator derived in Sec.~\ref{sec:3}.
The lowest DRPA solutions for both the $0^+$, $3^-$ and $1^-$ channels
were shifted down to their relative experimental energies and
then they were employed to generate the two-phonon contributions
for the ERPA calculation.
The spectrum obtained for  ${}^{16}{\rm O}$ is
displayed in Fig.~\ref{fig:2PiERPA_3g} and Table~\ref{tab:2PiERPA_3g}
together the total ph strength $Z_{n_\pi}$ of each state,
\begin{equation}
 Z_{n_\pi} ~=~  \sum_{\alpha \beta} ~
      \left| 
       {\mbox{$\langle {\Psi^A_{n_\pi} } \vert $}}
            c^{\dag}_\alpha c_\beta {\mbox{$\vert {\Psi^A_0} \rangle$}} 
      \right|^2   \; .
\label{eq:Z_tot_ph}
\end{equation}
 Table~\ref{tab:2PiERPA_3g} also
reposts the relative occupation of ph and two-phonon admixtures in
the wave function for each solution of the ERPA equation.


 The results yield an isoscalar $0^+$ 
state with a predominant ph character at $\sim$17~MeV,
as it was found in DRPA. Although this solution is now characterized
by a partial mixing with two-phonon configurations. 
 Table~\ref{tab:2PiERPA_3g} shows that this is the result
of mixing with 
the lowest solution in the same channel, which ends up at $\sim$11~MeV.
 The latter is predominately a two-phonon state.
  It is also seen that in both cases the
relevant configuration comes from the coupling of two $0^+$ phonons
themselves.
%
%
%
 The calculations of Ref.~\cite{iach1} have shown that the bulk of
the 4p4h contributions to the first excited state of ${}^{16}\textrm{O}$
may come from the coupling of four different phonons with negative
parity ($3^-$ and $1^-$). However,
the self-consistent role of  coupling positive parity states was
not considered in that work. Both this effect,
the RPA correlations and the inclusion of the nuclear fragmentation allow
for the --at least partial-- inclusion of configurations beyond the 2p2h case
even when no more than two-phonon coupling is considered, as in this paper.
 A study of the  importance of three- and four-phonon configurations
will be addressed in the future.

 The low-lying negative parity isoscalar states ($3^-$ and $1^-$) are
only slightly
 affected by two-phonon contributions and remain substantially
above the experimental energy at 9.23 and 10.90~MeV, respectively,
and more correlations will be needed in order to lower their energy.
We note that these wave functions contain a two-phonon admixture obtained by
coupling the low-lying $0^+$ excitation to either the $3^-$ or $1^-$ phonons.
 According to the above interpretation of the first $0^+$ excited state
this corresponds to the inclusion of 3p3h and beyond.
 At the same time two additional negative parity solutions, with two-phonon
character, are found at higher energy in accordance with experiment.
%

\begin{table}[t]
 \begin{center}
 \begin{tabular}{cccccccccccccccc}
  \hline 
  \hline 
 $T=0$
   & \hspace{.1truein}   & \multicolumn{2}{c}{dressed/DRPA} 
   & \hspace{.12truein}   & \multicolumn{5}{c}{dressed/ERPA}
       &  & $(0_2^+)^2$   & $(3_1^-)^2$
          & $(0_2^+,3_1^-)$ & $(0_2^+,1_1^-)$
          & \\
 $J^\pi$  &  & $\varepsilon^\pi_n$ & $Z_{n_\pi}$ 
          &  & $\varepsilon^\pi_n$ & $Z_{n_\pi}$
          & \hspace{.05truein} & $ph$(\%) & $2\Pi$(\%) 
      & ~ & (\%)   & (\%)  & (\%)   & (\%)
          & \\
  \hline 
  \hline 
 $1^-$ &  &        &         &   &  13.37 &  0.148 &   & 21   & 79   & &      &      &      &  79   & \\
 $3^-$ &  &        &         &   &  12.35 &  0.113 &   & 16   & 84   & &      &      & 84    \\
       \\
 $0^+$ &  &        &         &   &  12.15 &  0.001 &   &  1   & 99   & &  3   & 96   \\
 $4^+$ &  &        &         &   &  12.14 &  0.007 &   &  1   & 99   & &      & 99   \\
 $2^+$ &  &        &         &   &  12.12 &  0.008 &   &  1   & 99   & &      & 98   \\
       \\
 $0^+$ &  &  16.62 &   0.717 &   &  17.21 &  0.633 &   & 88   & 12   & &  10  &  0.5 \\
       \\
 $0^+$ &  &        &         &   &  11.28 &  0.092 &   & 12   & 88   & &  85  &  2   \\
       \\
 $1^-$ &  &  11.19 &   0.720 &   &  10.90 &  0.680 &   & 94.1 &  5.9 & &      &      &      &  5.8  & \\
 $3^-$ &  &   9.50 &   0.762 &   &   9.23 &  0.735 &   & 95.9 &  4.1 & &      &      &  4.0  \\
  \hline 
  \hline 
 \end{tabular}
  \end{center}
    \parbox[t]{1.\linewidth}{
      \caption[Two-phonon ERPA spectrum and strengths; dressed input]
  {  \small 
    Excitation energy and total spectral strengths obtained for the 
   principal solutions of DRPA and two-phonon ERPA equations.
    The total contribution of ph and two-phonon states
   of the ERPA solutions are shown.
    For states below 15~MeV, the columns on the right side give the 
individual contributions of
all the relevant two-phonon contributions.
    \label{tab:2PiERPA_3g}  }
    }
\end{table}

The two-phonon ERPA approach also generates a triplet of states at 
about 12~MeV, with quantum numbers $0^+$,  $2^+$ and $4^+$.
A similar triplet is found experimentally at 12.05, 11.52
and 11.10~MeV, which also corresponds to
twice the experimental energy of the first $3^-$ phonon.
 The ERPA solutions for this triplet are almost 
exclusively obtained by $3^- \otimes 3^-$ states and therefore have a 2p2h
character, in accordance
with Refs.~\cite{O16ph86,O162p2h}.
Further interaction in the pp and hh channels will be needed in order
to reproduce the correct experimental splitting and experimental
strength~\cite{iach1}.


\section{Conclusions}
\label{sec:concl}

 In the present paper the Faddeev technique, combined with self-consistent
 Green's Function theory,  has been applied to study 2h1p correlations 
at small missing energies for the nucleus of ${}^{16}{\rm O}$.
The application of the Faddeev method allows for the first time the treatment
of the coupling of ph and pp(hh) collective modes 
within a RPA framework, while the effects of fragmentation are included
through the dressing of the sp propagator.
The present approach allows to further identify the important role played
by the low-lying excited states of ${}^{16}{\rm O}$. These are seen
to be essential to generate many of the fragments with small spectrocopic
factors that are seen experimentally, like the $s_{1/2}$ and $p_{3/2}$
states of ${}^{15}{\rm N}$ at 5.20~MeV and $\sim$9~MeV. . 

The main impediment in obtaining a good theoretical description of
the single particle spectral function of ${}^{16}{\rm O}$ has been identified
in the poor description of the excitation spectrum, as obtained 
by solving the standard (D)RPA equations. We have improved on this by
computing the effects of mixing of ph states with two-phonon configurations
in the kernel of the Bethe-Salpeter equations.  Moreover, configurations
beyond the 2p2h level were also partially included by means
of improved RPA and self-consistency effects.
The results show that these contributions explain the formation
of several excited  states observed at low energy which are not 
obtained by RPA calculations.
 Still it appears that a full solution of the spectrum of ${}^{16}{\rm O}$
with this method requires to consider the mixing of up to 
four phonons and the interaction in the pp and hh channels~\cite{iach1}.
 The inclusion of such effects will be the topic of future work.

\acknowledgments
One of us (C.B.) acknowledges useful discussions with J.-M.~Sparenberg and
B.~Jennings.
This work was supported in part by the U.S. National Science Foundation
under Grants No.~PHY-9900713 and PHY-0140316 and in part by the Natural
Sciences and Engineering Research Council of Canada (NSERC).


\end{document}